# Equal Power Distribution and Dynamic Subcarrier Assignment in OFDM Using Minimum Channel Gain Flow with Robust Optimization Uncertain Demand

F.A. Hla Myo Tun, S.B. Aye Thandar Phyo, and T.C. Zaw Min Naing

**Abstract**— In this paper, the minimum channel gain flow with uncertainty in the demand vector is examined. The approach is based on a transformation of uncertainty in the demand vector to uncertainty in the gain vector. OFDM systems are known to overcome the impairment of the wireless channel by splitting the given system bandwidth into parallel sub-carriers, on which data-symbols can be transmitted simultaneously. This enables the possibility of enhancing the system's performance by deploying adaptive mechanisms, namely power distribution and dynamic sub-carrier assignments. The performances of maximumizing the minimum throughput have been analyzed by MATLAB codes.

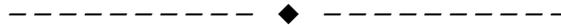

## 1 INTRODUCTION

THE **O**rthogonal **F**requency **D**ivision **M**ultiplexing (*OFDM*) appears as quite attractive transmission scheme for such future wireless systems: In an OFDM system, the total bandwidth is split into N parallel sub-channels, known as sub-carriers [1], where each sub-carrier can be assumed to experience flat fading. However, different sub-carriers of a broadband wireless system have a strongly varying attenuation, i.e. the system provides frequency diversity. This diversity can be exploited by dynamically adapting the modulation type and transmit power per sub-carrier at the transmitter, if the sub-carrier attenuations are known[2].

Today, many *dynamic sub-carrier assignment* algorithms have been proposed. Compared to static schemes these dynamic assignment algorithms can provide a performance increase of 100% per terminal, simply by utilizing the given bandwidth and transmit power much better. In the latter case each sub-carrier receives an equal amount of transmit power. Together with a target bit error probability the suitable modulation type can be obtained at once. In addition, the total transmit power is limited.

In this paper, Robust Optimization in Linear Programming was proposed for uncertain demand of minimum channel gain flow. The approach presented in this paper is based on a transformation of uncertainty in the supply/demand vector to uncertainty in the gain vector.

————————————————
- F.A. Dr. Hla Myo Tun is with the Mandalay Technological University, Mandalay, Myanmar.
- S.B. Ms. Aye Thandar Phyo is with the School of Computer Engineering, Nangyan Technological University, Singapore.
- T.C. Dr.Zaw Min Naing is with the Electronic Engineering Department, Technological University of Maubin, Maubin, Myanmar.

## 2 SYSTEM MODEL FOR DYNAMIC SUBCARRIER

### 2.1 Model of Wireless Channel

The terminals to be quasi-static and uniformly distributed over the cell are considered. Still, due to the movement of reflecting and scattering objects within the cell, the perceived signal quality per sub-carrier and terminal, i.e., their SNR varies permanently. The instant SNR of sub-carrier n for terminal j at time t is given by *SNR= (Power) × (Attenuation)/ (Noise Power)*. The attenuation is primarily responsible for the variation of the perceived SNR; it varies due to path loss, shadowing and fading. Thus, *Attenuation* can be decomposed into three factors reflecting these three effects. The attenuation of each sub-carrier is assumed to be constant over the time unit $T_f$. Note that this time unit is considered to be smaller than the coherence time of the wireless channel, using the definition from [4].

### 2.2 Problem Formulation and Objectives

We deal with a system that features dynamic sub-carrier assignment in combination with an equal power distribution per sub-carrier. As described in 2.1, each sub-carrier n is assigned to at most one terminal j at time t, indicated by a binary variable **x'** set to one (x' is set to zero if n is not assigned to j at t). Each sub-carrier is employed with equal transmit power. The adaptive modulation scheme is applied on top of the dynamic sub-carrier assignment and fixed power distribution. For the chosen objective, this results in an integer programming problem. Given a directed network G = (N, A) with finite node set N and finite arc set A, the minimum channel gain flow:

max     ∈

s.t     $\sum_j x' \leq 1$ for all n

$\sum_n F((\text{Power}) \times (\text{Attenuation})/(\text{Noise Power})).x' \geq \in$ for all j

where F((Power) × (Attenuation)/ (Noise Power)) describes the number of bits per downlink phase that can be transmitted on sub-carrier n for terminal j at time t with a transmit power. The first constraint guarantees the assignment of at most one terminal to one subcarrier at a time. The second constraint implements the discussed objective: Maximizing the minimum throughput per terminal per down-link phase. Note that no additional constraint for the transmit power is required as the power is statically distributed. In addition, notice the number of feasible solutions for this integer programming problem. Each sub-carrier can be assigned to one out of J terminals, thus a total of $J^N$ possible solutions exist.

## 3 PERFORMANCE ANALYSIS

In this section, we describe the methodology of Section 2. We also mention the tools we used to solve the integer programming problems with MATLAB environments. Finally, we present the results we derived from the simulations.

### 3.1 Methodology

In order to obtain the system level results, we took the following steps: Initially we generated channel trace files of the attenuation values for each sub-carrier regarding each terminal. One sample was generated for every down-link phase and sub-carrier per terminal. Theses attenuation values were used to generate and solve appropriate instances of the linear programs described in Secton 2 (using the MATLAB [2]).

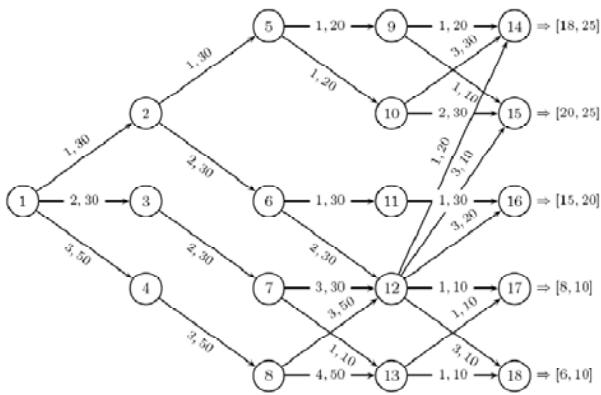

Fig.1. Example Network for Uncertain Demand for Power Adaption

### 3.2 FUNCTION Creation in MATLAB

The MATLAB function of maximizing the minimum througnput for power distribution and dynamic subcarrier assignment has two main parts. The SNR and minimum_throughput are calculated by linear integer programming equations. Finally, the maximum function for minimum_throughput can be evaluated by numerical calculation method. The channel allocation results for channel gain can be got by linprog function. We have to apply the example network from Fig.1 to test the result of uncertain demand network.

### 3.3 Experimental Results

After setup the creation of MATLAB functions in workspace, the maximizing the minimum throughput for uncertain demand network flow result will be appeared as Fig.2. The SNR depends on the transmit power results shown in Fig.3 are evaluated by the maximum gain network function.

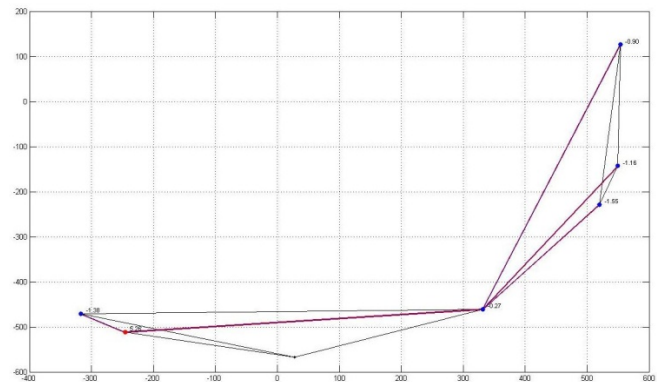

Fig.2. Maximizing the Minimum Throughput for Uncertain Demand Network Flow Result

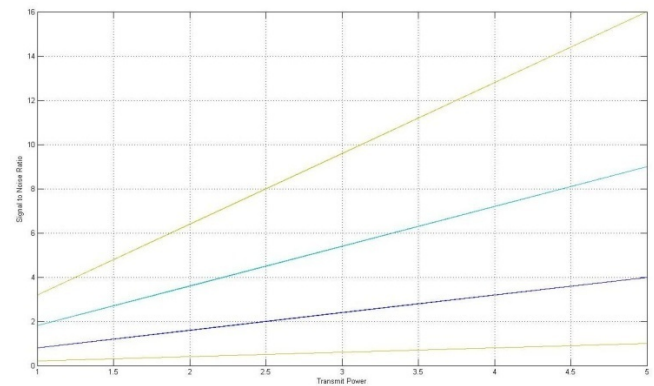

Fig.3. Signal to Noise Ratio with respect to Transmit Power

## 4 CONCLUSION

In this paper, we analyzed the most popular adaptive mechanisms that are used to enhance OFDM-FDMA systems, *power adaption* and *dynamic sub-carrier assignment*, in terms of the achieved minimum throughput. If the transmit power is low or the attenuation spread in the cell is high we propose to consider a fully dynamic approach due to the quite high additional peformance achieved. The results from MATLAB code have been implemented with respective functions for uncertain demand.

## ACKNOWLEDGMENT

I would wish to acknowledge the many colleagues at Mandalay Technological University who have contributed to the development of this paper. In particular, I would like to thank Myat Su Nwe, my wife, Thet Htar Swe, my daughter, and Zay Yar Tun, my son, for their complete support.

## REFERENCES

[1] R. van Nee and R. Prasad,, "*OFDM Wireless Multimedia Communications*," chapter 9, Artech House, 2000.

[2] Mathias Bohge, James Gross, Adam Wolisz, *The Potential of Dynamic Power and Sub-carrier Assignments in Multi-User OFDM-FDMA Cells.* In Proc. of IEEE Globecom 2005, St. Louis, MO, USA, November 2005.

[3] Simone Gast, "Applying Robust Optimization to Network Flow Problems with uncertain demand," *Department of Mathematics, University of Erlangen-N¨urnberg, 2008.*

[4] R.K. Ahuja, T.L. Magnanti, and J.B. Orlin., "Network Flows: Theory, Algorithms, and Applications," Prentice Hall, 1993.

[5] D. Kivanc, G. Li, and H. Liu, "*Computationally efficient bandwidth allocation and power control for OFDMA*," IEEE Transactions on Wireless Communications, vol. 2, no. 6, pp. 1150–1158, 2003.

[6] M. Bohge, "Bit loading versus dynamic sub-carrier assignment in multiuser OFDM-FDMA systems," September 2004, Diploma Thesis at Technical University of Berlin, Germany.

**First A.** Dr. Hla Myo Tun received his Bachelor Degree in Electronic from Mandalay Technological University, Mandalay in 2002, his Master Degree in Electronic from Yangon Technological University, Yangon in 2003 and his Doctoral Degree in Electronic from Mandalay Technological University, Mandalay in 2008. He has published eleven conference papers and three journal publications. His current research interests are Real Time Distributed Control System in Industrial Automation, Missile Control Using MATLAB. Dr. Tun is a member in Myanmar Engineering Society (MES). He received one papers from IEEE publications in 2008.
.
**Second B. Ms. Aye Thandar Phyo** received her bechlor degree from Technological University (Hmawbi) in 2006. She is now doing research in NTU, Singapore with Network Flow in uncertain demand.

**Third C. Dr. Zaw Min Naing** is a member of the MES and the Pro Rector of Technological University (Maubin).